\documentclass[conference]{IEEEtran}
\IEEEoverridecommandlockouts
\usepackage{cite}
\usepackage{amsmath,amssymb,amsfonts}
\usepackage{algorithmic}
\usepackage{textcomp}
\renewcommand\IEEEkeywordsname{Keywords}
\usepackage{graphicx}
\usepackage{float}
\usepackage{listings}
\floatstyle{plaintop}
\restylefloat{table}
\def\BibTeX{{\rm B\kern-.05em{\sc i\kern-.025em b}\kern-.08em
    T\kern-.1667em\lower.7ex\hbox{E}\kern-.125emX}}

\usepackage{geometry}
\makeatletter
\newcommand{\newlineauthors}{%
  \end{@IEEEauthorhalign}\hfill\mbox{}\par
  \mbox{}\hfill\begin{@IEEEauthorhalign}
}
\makeatother

\begin{document}

\title{Accelerating GPU-Based Out-of-Core Stencil Computation with On-the-Fly Compression}

\author{
\IEEEauthorblockN{Jingcheng Shen}
\IEEEauthorblockA{\textit{Graduate School of Information}\\
\textit{Science and Technology} \\
\textit{Osaka University} \\
Osaka 565-0871, Japan \\
jc-shen@ist.osaka-u.ac.jp}
\and
\IEEEauthorblockN{Yifan Wu}
\IEEEauthorblockA{\textit{Graduate School of Information}\\
\textit{Science and Technology} \\
\textit{Osaka University} \\
Osaka 565-0871, Japan \\
yf-wu@ist.osaka-u.ac.jp}
\newlineauthors
\IEEEauthorblockN{Masao Okita}
\IEEEauthorblockA{\textit{Graduate School of Information}\\
\textit{Science and Technology} \\
\textit{Osaka University} \\
Osaka 565-0871, Japan \\
okita@ist.osaka-u.ac.jp}  
\and
\IEEEauthorblockN{Fumihiko Ino}
\IEEEauthorblockA{\textit{Graduate School of Information}\\
\textit{Science and Technology} \\
\textit{Osaka University} \\
Osaka 565-0871, Japan \\
ino@ist.osaka-u.ac.jp}  
}
\maketitle

\begin{abstract}
Stencil computation is an important class of scientific applications that can be efficiently executed by graphics processing units (GPUs). Out-of-core approach helps run large scale stencil codes that process data with sizes larger than the limited capacity of GPU memory.
However, the performance of the GPU-based out-of-core stencil computation is always limited by the data transfer between the CPU and GPU. Many optimizations have been explored to reduce such data transfer, but the study on the use of on-the-fly compression techniques is far from sufficient. 
In this study, we propose a method that accelerates the GPU-based out-of-core stencil computation with on-the-fly compression. We introduce a novel data compression approach that solves the data dependency between two contiguous decomposed data blocks. We also modify a widely used GPU-based compression library to support pipelining that overlaps CPU/GPU data transfer with GPU computation. Experimental results show that the
proposed method achieved a speedup of 1.2$\times$ compared the method without compression. Moreover, although the precision loss involved by compression increased with the number of time steps, the precision loss was trivial up to 4,320 time steps, demonstrating the usefulness of the
proposed method.            
\end{abstract}

\section{Introduction}
\label{sec:intro}
Stencil computation is an important computation paradigm that appears in many scientific applications, such as geophysics simulations~\cite{serpa17padw,farres19eage,shen20ieice}, computational electromagnetics~\cite{adams07fdtd}, and image processing~\cite{tabik18supe}. The key principle of stencil computation is to iteratively apply a fixed calculation pattern (stencil, where the update of an element relies on the surrounding elements) to every element of the output datasets.
Such a single-instruction multiple-data (SIMD) characteristic of stencil computation makes itself a perfect scenario to use the graphics processing units (GPUs) for acceleration.
A GPU has thousands of cores and its memory bandwidth is 5--10 times higher than that of a CPU, thus excelling at accelerating both compute- and memory-intensive scientific applications~\cite{okuyama14itpds,ikeda14ieeejbhi,shen17ica3pp,shen19ccpe}. 
However, as a GPU has a limited capacity of device memory (tens of GBs), it fails to run a large stencil code directly whose data size exceeds its memory capacity.

A large entity of research on GPU-based out-of-core stencil computation has been performed to address this issue~\cite{jin14ichpca,sourouri16ijpp,shimokawabe17iccc,miki19hpcn,shen20ieice}. For a large dataset whose data size exceeds the capacity of the device memory, out-of-core computation first decomposes the dataset into smaller blocks and then streams the blocks to and from the GPU for processing. Nevertheless, the outcome of this approach is often limited by data transfer between the CPU and GPU because the interconnects fail to catch up with the development of the computation capability of GPUs as described in \cite{shen19ccpe}. Data-centric strategies are thus necessary to reduce the data transfer between the CPU and GPU.
Studies have introduced strategies such as temporal blocking and region sharing to reuse the on-GPU data and to avoid transferring extra data \cite{jin14ichpca,miki19hpcn,shen20ieice}.
Nevertheless, we need to further optimize the methods to reduce data transfer time. according to \cite{shen20ieice}, the performance of out-of-core stencil code is still limited by data transfer.
A potential solution is to use on-the-fly compression to compress the data on the GPU before transferring back to the CPU, and decompress the data on the GPU before processing. Until now, however, studies on the acceleration of GPU-based out-of-core stencil computation with on-the-fly compression are really rare. According to a comprehensive review \cite{cappello20review}, research on leveraging lossy compression techniques in scientific applications mainly focuses on scenarios such as post-analysis and failure recovery. 
We think that the scarcity of relevant research raises two research questions: 
\begin{itemize}
    \item  Would the overhead of compression/decompression outweighs the reduced data transfer time?
    \item  Would the precision loss involved by data compression be so huge that the output becomes useless?
\end{itemize}

In this study, we (1) propose a method to accelerate out-of-core stencil computation with on-the-fly compression on the GPU and (2) try to give answers to the two above-mentioned questions. The contribution of this work is three-fold: 
\begin{itemize}
    \item We introduced a novel approach to integrate an on-the-fly lossy compression into the workflow of a 25-point stencil computation. For large datasets that are decomposed into blocks, this approach solves the data dependency between two contiguous blocks and thus secures the accessibility to the common regions between two contiguous blocks after compression.      
    \item We modified a widely-used GPU-based compression library~\cite{cuZfp} to support pipelining, which is mandatory for the purpose of overlapping CPU-GPU data transfer with GPU computation.
    \item We gathered experimental results to answer the aforementioned questions, i.e., on-the-fly compression is useful in reducing the overall execution time of out-of-core stencil computation, and the precision loss is tolerable. Therefore, the present study may lead to future research on leveraging compression techniques to accelerate out-of-core stencil computation on a GPU.  
\end{itemize}

The remainder of this study is organized as follows: Previous studies that accelerate stencil and similar scientific applications with compression techniques are introduced in Section~\ref{sec:prew}. Stencil computation, its background and challenges in the acceleration of stencil computation with on-the-fly compression are briefly described in Section~\ref{sec:stencil}. Section~\ref{sec:compr} discusses the selection of an appropriate GPU-based compression library. The proposed method used to integrate the compression processes into the workflow of out-of-core stencil computation is described in Section~\ref{sec:proposed}. In Section~\ref{sec:expr}, experimental results are presented and analyzed. Finally, Section~\ref{sec:conc} concludes the present study and proposes future research directions.

\section{Previous Work}
\label{sec:prew}
Nagayasu \textit{et al.}~\cite{nagayasu08cg} proposed a decompression pipeline for accelerating out-of-core volume rendering of time-varying data. Their method was specified to compress and decompress RGB data and the decompression procedure was partially performed on the CPU.  

Tao \textit{et al.}~\cite{tao18hpdc} proposed a lossy checkpointing scheme, which significantly improved the checkpointing performance of iterative methods with lossy compressors. In the presence of system failures, their method reduced the fault tolerance overhead for iterative methods by 23\%--70\% compared with traditional checkpointing and 20\%--58\% compared with lossless-compressed checkpointing. 

Calhoun \textit{et al.}~\cite{calhoun19saga} proposed metrics to evaluate loss of accuracy caused by using lossy compression to reduce the snapshot data used for checkpoint restart. In this study, Calhoun and colleagues improved efficiency in checkpoint restart for partial differential equation (PDE) simulations by compressing the snapshot data, and found that this compression did not affect overall accuracy in the simulation, as demonstrated by the proposed evaluation metrics.

Wu \textit{et al.}~\cite{wu19sc} proposed a method to simulate large quantum circuits using lossy or lossless compression techniques adaptively. Thanks to their method, they managed to increase the simulation size by 2--16 qubits. However, their method was designed for CPU-based supercomputers and thus the compression libraries cannot be used for GPU-based scenarios. Moreover, the adaptive selection between lossy and lossless compression, i.e., using lossy compression if lossless one failed, is impractical in GPU-based applications because such failures heavily impair the computational performance.  

Jin \textit{et al.}~\cite{jin20ipdps} proposed a method to use GPU-based lossy compression for extreme-scale cosmological simulations. Their findings show that GPU-based lossy
compression can enable sufficient accuracy on post-analysis for cosmological simulations, as well as high compression and decompression throughputs. Instead of compressing datasets for post-analysis, our method aims to improve the performance of GPU-based out-of-core stencil computation by compressing or/and decompressing datasets at run-time.

Tian \textit{et al.}~\cite{tian20pact} proposed Cusz, an efficient GPU-based error-bounded lossy compression framework for scientific computing. This framework reported high compression and decompression throughputs and a good compression ratio. However, according to their study, Cusz has sequential subprocedures, which prevents us to use this framework as on-the-fly compression in our work due to the concern of the overhead to shift from GPU to CPU computation. 

Zhou \textit{et al.}~\cite{zhou21ipdps} designed high-performance MPI libraries with on-the-fly compression for modern GPU clusters. In their work, they reduced the inter-node communication time by compressing the messages transferred between nodes, and the size of messages was up to 32 MB. On the other hand, our method compressed large datasets for stencil computation that were more than 10 GB to reduce the data transfer time between the CPU and GPU (i.e., intra-node communication time). Moreover, our method is specified to deal with out-of-core stencil computation, solving the data dependency between decomposed data blocks.

\section{Stencil Computation}
\label{sec:stencil}
\begin{figure}
    \centering
    \includegraphics[width=\linewidth]{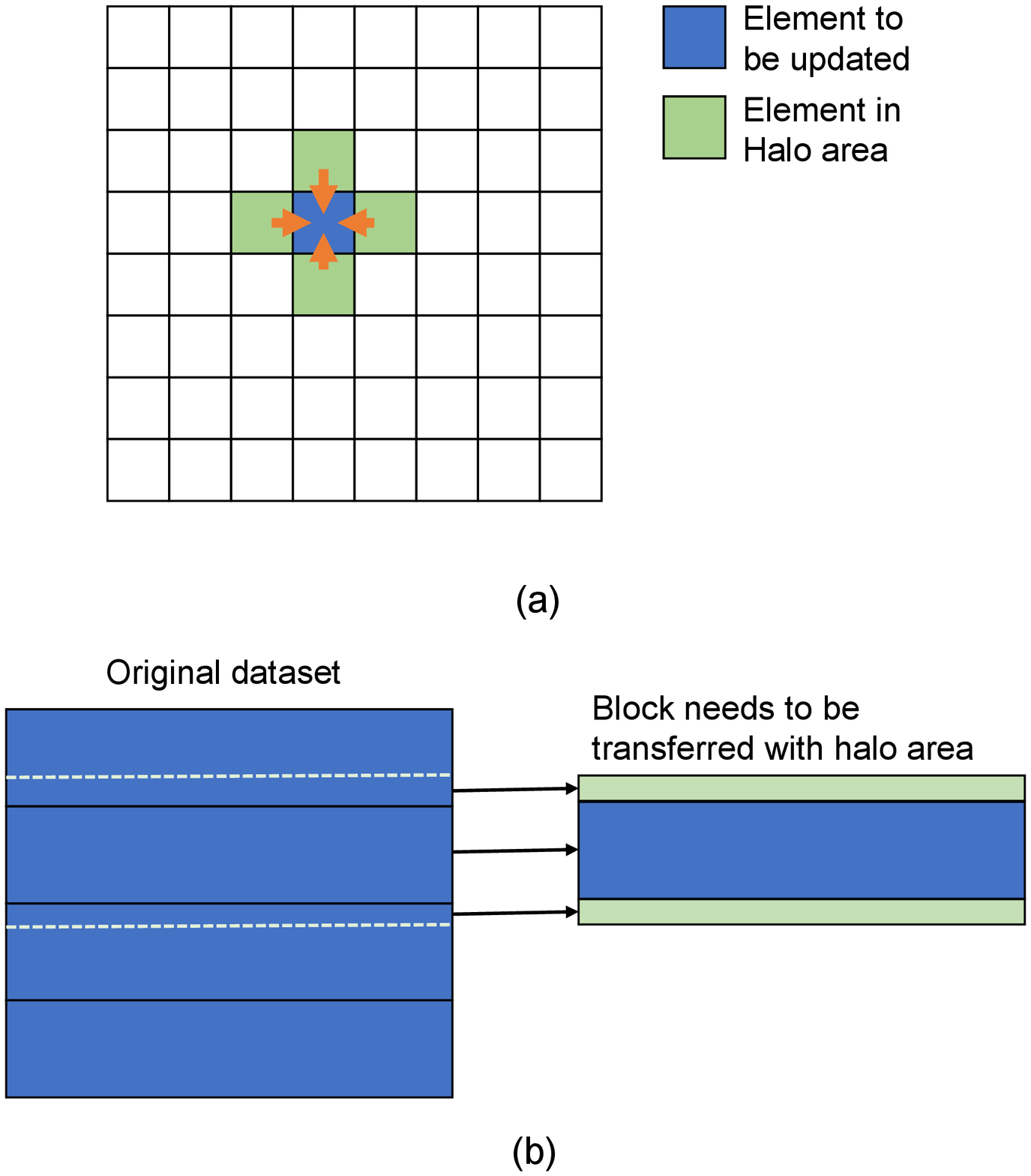}
    \caption{Five-point stencil computation. Subfigure (a) update of an element relies on its four neighbor elements. Subfigure (b) the decomposed blocks must be transferred with the according halo data.}
    \label{fig:stencil}
\end{figure}
\begin{figure}
    \centering
    \includegraphics[width=\linewidth]{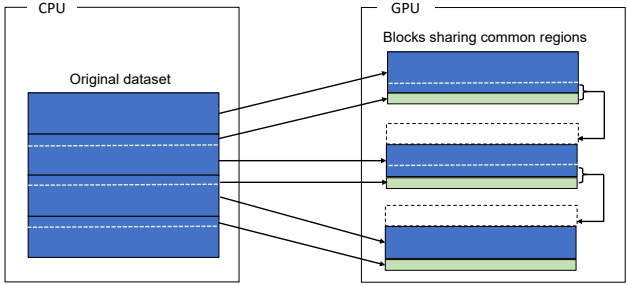}
    \caption{The contiguous blocks can share common regions on the GPU
and thus avoid transferring the amount of data that is equivalent to the
halo areas.}
    \label{fig:shareregions}
\end{figure}

Stencil computation is an iterative computation that updates each element in one or more datasets according to a fixed pattern that computes with the neighbor elements of the element to be updated. 
A hello-world application of stencil computation is the solver of Laplace's equation, which can describe the phenomenon of heat conduction: A five-point stencil code, where the temperature of each element at the (\textit{t}+1)-th time step is obtained by taking the average temperature of the four surrounding elements at the \textit{t}-th time step (Fig. \ref{fig:stencil}(a)).

To use out-of-core stencil computation, we decompose the large datasets into smaller blocks and stream the blocks to and from the GPU for processing. Because of the data dependency of stencil computation, when we transfer a block to the GPU for computation, we must also piggyback the neighbor data (``halo area'' in jargon) with the block (Fig. \ref{fig:stencil}(b)). The more time steps we want to compute on a block on the GPU, the larger halo area we must transfer along with the block. But because two contiguous blocks share common regions, a block can get
the common regions from its former block and provide the common regions with its later block. By doing so, we can in effect reduce the amount of data transfer equivalent to the size of halo areas (Fig. \ref{fig:shareregions}).

One challenge in integrating on-the-fly compression into the workflow of out-of-core stencil computation is that we must solve the aforementioned data dependency. Naively compressing each block prevents the use of halo
areas, whereas compressing each block with its halo areas not only consumes more memory space but also prevents contiguous blocks sharing common regions. Therefore, sophisticated compression strategy is necessary, and will be introduced in Section \ref{sec:sep}.

\section{On-the-fly Compression}
\label{sec:compr}
Another concern in using on-the-fly compression in out-of-core computation is the overheads of compression and decompression that are often considerable. GPU-based compression libraries such as cuZFP \cite{cuZfp}, Cusz \cite{tian20pact}, and nvComp \cite{nvcomp} reported high speeds in compression and decompression. The cuZFP and Cusz libraries are lossy compression, whereas the nvComp is lossless.

In this study, we used cuZFP because this library is performant and its source code is (relatively) easy to modify to implement functionalities we want. The cuZFP library allows users to specify the compression ratio.
Users can specify the number of bits to use to preserve a value. For example, specifying 32 bits to preserve a double-precision floating-point (i.e., double-type) value achieves a compression ratio of 2:1.

We avoid using the lossless nvComp due to the concern of compression ratio. In the preliminary experiments, we found the size of data compressed with nvComp was larger than that of the uncompressed data. We therefore avoid the use of nvComp because we cannot estimate the upper bound of the size of the compressed data, so we must allocate device memory every time the compression happens instead of reusing pre-allocated device buffers with fixed sizes. The reason why we avoided using Cusz was explained in Section \ref{sec:prew}.

\section{Proposed Method}
\label{sec:proposed}
In this section, we introduce our proposed method, including separate compression that solves the data dependency between contiguous blocks and thus allows us to compress the decomposed datasets freely, and a pipelining version of cuZFP that supports overlapping compression/decompression with CPU-GPU data transfer.
\subsection{Separate Compression}
\label{sec:sep}
\begin{figure}
    \centering
    \includegraphics[width=\linewidth]{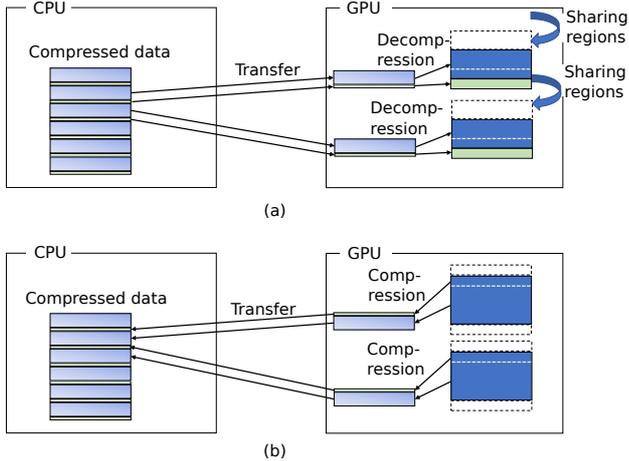}
    \caption{Separate compression approach to solve data dependency between contiguous blocks. In this approach, a compressed block consists of the reminder and the common region. As shown in subfigure (a), the $i$-th compressed reminder and common region are decompressed on the GPU for computation; and in subfigure (b), after computation, the reminder and common region were compressed and transferred back to CPU to update the $i$-th remainder and ($i-1$)-th common region, respectively.}
    \label{fig:sepcomp}
\end{figure}
As shown in Fig. \ref{fig:shareregions}, two contiguous blocks have
common regions that are shareable. Precisely, the bottom
halo areas needed by the $i$-th block fall in the ($i+1$)-th
block, and the top halo areas needed by the ($i+1$)-th
block fall in the the ($i$)-th block. Therefore, the common
regions between the two blocks consist of the top areas
and a part of the ($i+1$)-th block whose size is equivalent
to that of the top halo areas. If we transfer the ($i$)-th block
with its bottom halo areas, we can avoid transferring the
common regions for the ($i+1$)-th block.

Summarily, each block only needs to be transferred
with its reminder and bottom halo areas, so the two
parts, i.e., reminder and half common region, must be
exclusively readable and writable to the according contiguous
blocks. Based on the observation, we propose
a separate compression approach that compress the two
parts separately. As shown in Fig. \ref{fig:sepcomp}(a), before computation, the $i$-th compressed reminder and common region
are decompressed and therefore the $i$-th block can be
computed on and provide the data needed by the $(i+1)$-
th block. As shown in Fig. \ref{fig:sepcomp}(b), after computation, the
$(i+1)$-th block are compressed as the $(i+1)$-th reminder
and  $i$-th common region.
\subsection{Pipelining cuZFP}
\label{sec:pzfp}
\begin{figure}
    \centering
    \includegraphics[width=\linewidth]{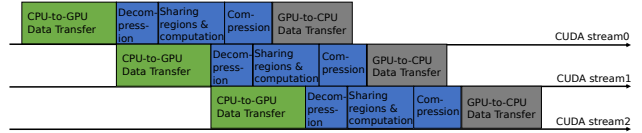}
    \caption{Modified cuZFP that supports pipelining. Three CUDA streams are used to perform operations, overlapping CPU-GPU data transfer with GPU kernels including compression, decompression, and computation.}
    \label{fig:pzfp}
\end{figure}
The cuZFP library \cite{cuZfp} is mainly designed as a standalone
tool that can be seamlessly used for post-analysis
and CPU-centric scientific computations. However, as an
on-the-fly process in the out-of-core stencil computation,
we must modify the source code to support pipelining
that overlaps CPU/GPU data transfer with GPU computations.
Thanks to the excellent coding quality of the
cuZFP project, we managed to modify the source code
to add such functionality with a reasonable amount of
programming effort. In the pipelining cuZFP, we use three
CUDA \cite{cudawww} streams to perform operations (Fig. \ref{fig:pzfp}).
\section{Experimental Results}
\label{sec:expr}
\begin{table}[tb!]
    \centering
    \begin{tabular}{lllll} \\\hline
        No. of & Data type & Dim. info. & Entire   \\
        datasets &  &  & data size   \\\hline
        4 & Double & (1152+2$\times$HALO)$^3$ &  46 GB  \\
          &  & HALO$=$4 &  \\\hline
    \end{tabular}
    \caption{Target stencil code.}
    \label{tab:dataset}
\end{table}
\begin{table}[tb!]
    \centering
    \begin{tabular}{ll} \\\hline
        GPU    &  NVIDIA Tesla V100-PCIe\\
        Device memory &  32 GB\\
        CPU    &  Xeon Silver 4110 \\
        Host memory &  500 GB \\
        OS     &  Ubuntu 16.04.6 \\
        CUDA & 10.1 \\
        cuZFP & 0.5.5\\\hline
    \end{tabular}
    \caption{Testbed for experiments.}
    \label{tab:machine}
\end{table}

In this section, we analyze the experimental results to evaluate the benefits of using on-the-fly compression in out-of-core stencil computation on a GPU. The stencil code we used is an acoustic wave propagator from a previous work \cite{shen20ieice} of ours. The code is a 25-point stencil computation that has two read-write datasets, a write-only dataset, and a read-only dataset. The two read-write datasets store the updated elements of the volume, and needed to be transferred to and from the GPU. The write-only dataset stores intermediate results at run-time and does not need to be transferred at all. The read-only dataset are constant values that must be referenced at run-time, and needed to be transferred to the GPU. The values are of double-type because it is more preferable compared to single-precision floating-point format (i.e. \texttt{float}-type) in iterative scientific applications. According to our previous work~\cite{shen20iccc}, the CPU version of a code using float-type data leads to outputs different from that of the GPU version. Such divergence becomes a more severe problem with the increase of the total number of iterations. On the other hand, when using double-type, results of the CPU and GPU versions of the same code  were consistent. Table \ref{tab:dataset} shows the detail of the datasets used by the stencil code. 

Moreover, we used four codes in our experiments to evaluate the performance and precision loss. The four codes include:
\begin{enumerate}
   \item The original stencil code.
    \item The stencil code with one read-write dataset compressed using a 32/64 rate (i.e., using 32 bits to preserve each double value).
    \item The stencil code with the read-only dataset compressed using a 32/64 rate.
    \item The stencil code with one read-write dataset and the read-only dataset compressed using a 24/64 rate. Note that we used 24 bits to preserve each double value due to the limited device memory capacity.
\end{enumerate}
The configuration to run the stencil codes is as the one described in \cite{shen20ieice} where the number of division is 8 and the number of temporal blocking time steps is 12. Accordingly, we divide the data into 8 blocks, and when a block is transferred to the GPU, it will be computed on for 12 times before transferred back to the CPU. For the total time steps, we used numbers from 480 to 4320 with an increment of 480. For specifications of the testbed for all experiments performed, see Table~\ref{tab:machine}.

\subsection{Evaluation of Performance Benefits}
\label{sec:benef}
\begin{figure}
    \centering
    \includegraphics[width=\linewidth]{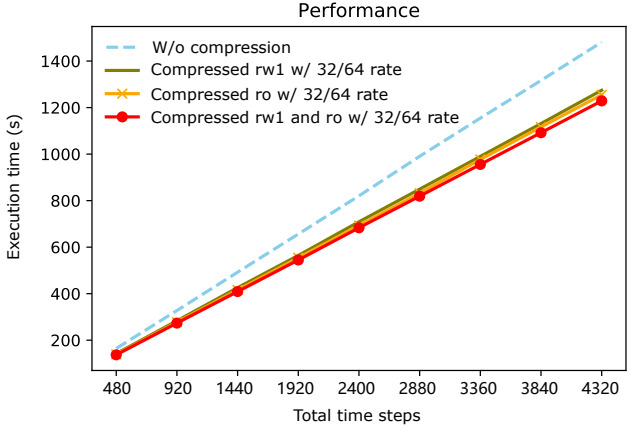}
    \caption{Performance of the four stencil codes.}
    \label{fig:spdup}
\end{figure}
As shown in Fig. \ref{fig:spdup}, the three codes using on-the-fly compression ran faster than the original code. The code compressing a read-write dataset and a read-only dataset outperformed the others, running 1.20$\times$ as fast as the original code. The code compressing a read-only dataset and the code compressing a read-write dataset achieved speedups of 1.18$\times$ and 1.16$\times$, respectively. Based on these results, our proposed method is beneficial for GPU-based out-of-core stencil computation in terms of performance. A detailed analysis of the achieved performance improvement will be given in next section. 
\subsection{Detailed Analysis of Achieved Performance Improvement}
\label{sec:detailed}
\begin{figure}
    \centering
    \includegraphics[width=\linewidth]{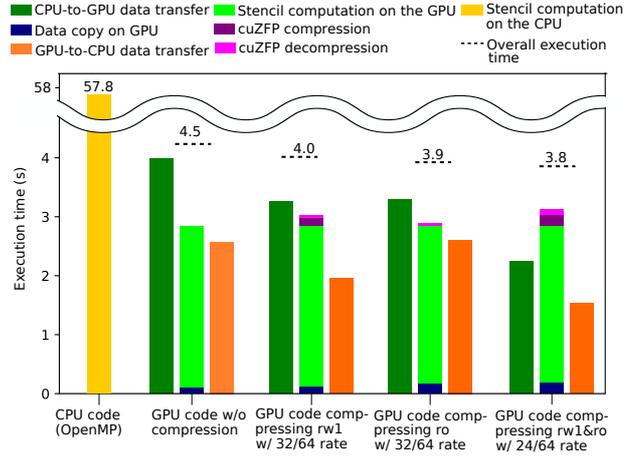}
    \caption{Breakdown of the execution time when the four GPU-based codes ran for 12 time steps. The execution time of a CPU-based code was measured to show the performance benefits of using GPU acceleration. Note that the bounding operation time for the fourth GPU-based code was GPU computation time (bars in the middle), whereas the bounding operation time for the other three GPU-based codes was CPU-to-GPU data transfer time (dark green bars).}
    \label{fig:detail}
\end{figure}
In this experiment, we ran the four GPU-based codes individually for 12 time steps and profiled the breakdown of execution time. Moreover, we also ran a CPU-based code for 12 time steps to show the advanced performance of GPU-based code, compared to that of the CPU-based code. The CPU-based code was parallelized with OpenMP~\cite{wwwomp} and executed with 40 CPU threads. As shown in Fig. \ref{fig:detail}, we can see the three codes using compression reduced the CPU-to-GPU time (dark green bars) that limited the overall performance. The most interesting finding is that the fourth GPU-based code shifted from data-transfer-bounding to computation-bounding compared to the former three GPU-based codes, which is favorable because it theoretically means that the data transfer time can be fully hidden by the computation time. 

Moreover, although the code compressing the read-only dataset did not reduce the GPU-to-CPU data transfer time, it did not involve significant compression time (dark purple). Therefore, the code compressing the read-only dataset slightly outperformed the code compressing a read-write dataset. Nevertheless, the gaps between the overall execution time and the bounding operation time (i.e., longest bar) of the three codes with compression are larger than that of the original GPU-based code. This suggests that the compression or/and decompression involved some unidentified overheads that compromised the efficiency of overlapping data transfer with GPU computation, otherwise the overall execution time should have been similar to the bounding operation time. Therefore, we need more sophisticated measures to orchestrate the pipelining to achieve further improvement.  
\subsection{Evaluation of Precision Loss}
\label{sec:mod}
\begin{figure}
    \centering
    \includegraphics[width=\linewidth]{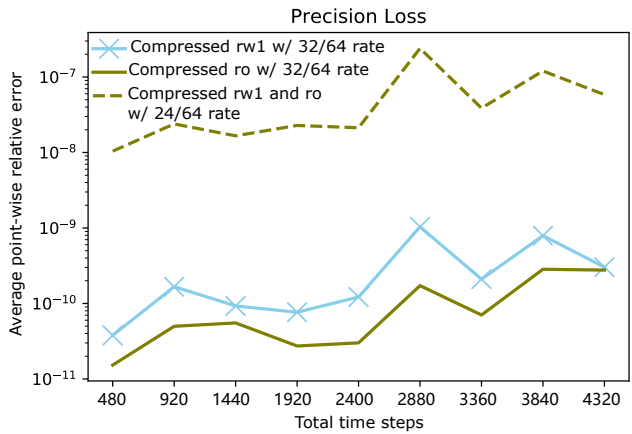}
    \caption{Change in precision loss as total time steps increase.}
    \label{fig:precloss}
\end{figure}
Besides showing performance benefits, demonstrating that the compression involves no significant precision loss is crucial. After completing the total time steps, we sampled 115,200 points (i.e., 100 points per plane) and compared the point values of the three codes using compression with that of the original code to calculate the average point-wise relative errors (Fig. \ref{fig:precloss}). Although the relative errors increased with an increase in the total time steps, they were still far from significant at 4,320 time steps. The code compressing the read-only dataset had the lowest precision loss because the read-only dataset does not need to be compressed repeatedly. The code compressing one read-write dataset and the read-only dataset using 24/64 rate resulted in the largest precision loss due to the fewer bits we used to preserve the double values. Nevertheless, the code is useful because the relative error was trivial (between $10^{-6}$ and $10^{-7}$). Given this, the proposed method will not lead to intolerable precision loss at least for a moderate number of time steps.   

\section{Conclusions and Future Work}
\label{sec:conc}
In this study, we introduced a method to accelerate GPU-based out-of-core stencil computation with on-the-fly compression. To realize the method, we proposed a novel approach to compress the decomposed data, solving the data dependency between contiguous blocks. We also modified the cuZFP library \cite{cuZfp} to support pipelining for overlapping data transfer with GPU computation. Experimental results show that the proposed method achieved a speedup of 1.2$\times$ at the expense of a trivial precision loss, i.e., an average point-wise relative error between $10^{-6}$ and $10^{-7}$. 
The results answer the two research questions mentioned in Section \ref{sec:intro}. First, the reduction of CPU-GPU data transfer time achieved by using on-the-fly compression outweighs the overhead of compression/decompression, improving the overall performance of GPU-based out-of-core stencil computation.  
Secondly, the on-the-fly compression does not cause severe precision loss for thousands of time steps.
Future work includes (1) comparing other on-the-fly compression algorithms to cuZFP and (2) orchestrating the pipelining for better efficiency in overlapping data transfer with GPU computation.

\section*{Acknowledgments}
This study was supported in part by the Japan Society for the
Promotion of Science KAKENHI under grants 20K21794, and “Program for Leading Graduate
Schools” of the Ministry of Education, Culture, Sports, Science,
and Technology, Japan.
We collaborated with Mauricio Hanzich and Albert Farr\'es (Computer Applications in Science and Engineering (CASE) Department, Barcelona Supercomputing Center (BSC)) to implement the stencil code used in this work.
%
%

\end{document}